\documentclass[iop,apjl,tighten,revtex4]{emulateapj}
\usepackage{color}
\usepackage{graphicx}
\usepackage{natbib}
\usepackage{amssymb}
\usepackage{amsmath}
\usepackage{rotating}
\usepackage{url}

\shorttitle{A Luminous Red Nova in M31 and its Progenitor System}
\shortauthors{Williams et~al.}

\begin{document}

\title{A Luminous Red Nova in M31 and its Progenitor System}

\author{S.~C. Williams, M.~J. Darnley, M.~F. Bode and I.~A. Steele}
\affil{{\scriptsize Astrophysics Research Institute, Liverpool John Moores University, IC2 Liverpool Science Park, Liverpool, L3~5RF, UK; scw@astro.ljmu.ac.uk}}
\submitted{{\scriptsize Received 2015 March 4; accepted 2015 April 28}}
\journalinfo{The Astrophysical Journal Letters, Draft version \today}

\begin{abstract}
We present observations of M31LRN~2015 (MASTER OT J004207.99+405501.1), discovered in M31 in 2015 January, and identified as a rare and enigmatic luminous red nova (LRN).  Spectroscopic and photometric observations obtained by the Liverpool Telescope showed the LRN becoming extremely red as it faded from its $M_V=-9.4\pm0.2$ peak. Early spectra showed strong H$\alpha$ emission that weakened over time as a number of absorption features appeared, including Na~{\sc i} D and Ba~{\sc ii}. At later times strong TiO absorption bands were also seen. A search of archival {\it Hubble Space Telescope} data revealed a luminous red source to be the likely progenitor system, with pre-outburst H$\alpha$ emission also detected in ground-based data. The outburst of M31LRN~2015 shows many similarities, both spectroscopically and photometrically, with that of V838 Mon, the best studied LRN. We finally discuss the possible progenitor scenarios.
\end{abstract}

\keywords{binaries: general --- novae, cataclysmic variables --- stars: individual (M31LRN~2015, V838 Mon) --- stars: peculiar --- supergiants}

\section{Introduction}
Luminous red novae (LRNe) are a rare type of stellar transient typically more luminous than most novae, but fainter than supernovae (SNe), and characterized by a reddening color as they fade. The canonical system is V838 Monocerotis, which produced an outburst in early 2002, peaking at $M_V=-9.95$ \citep{2002A&A...389L..51M} and later generating a spectacular light echo \citep{2003Natur.422..405B}. The only extragalactic red transient with a similar peak magnitude is M31-RV \citep{1989ApJ...341L..51R}. Additionally, several other extragalactic transients show some similar properties to LRNe, but are significantly more luminous at peak (e.g., M85~OT2006-1, \citealp{2007Natur.447..458K}; PTF~10fqs, \citealp{2011ApJ...730..134K}), and often grouped as intermediate luminosity red transients (ILRTs).

A number of mechanisms have been proposed to explain these events, including (1)~a classical nova mechanism \citep{2010ApJ...725..831S}, (2)~an unusual SN mechanism (e.g.\ \citealp{2013ApJ...769..109L}), (3) giant planet capture \citep{2003MNRAS.345L..25R}, (4)~stellar mergers \citep{2003ApJ...582L.105S} and (5)~extreme AGB stars \citep{2009ApJ...705.1364T}. \citet{2011ApJ...741...37K}, in a study of ILRTs SN 2008S and NGC 300 OT2008, suggested that LRNe and ILRTs may have different physical origins, favoring the extreme AGB scenario as the likely mechanism for ILRTs. \citet{2011A&A...528A.114T} presented compelling evidence that V1309 Scorpii, a Galactic LRN, was a merger of a contact binary.

MASTER OT J004207.99+405501.1 (hereafter M31LRN~2015) was discovered on 2015 January 13.63~UT by the MASTER-Kislovodsk auto-detection system at $00^{\mathrm{h}}42^{\mathrm{m}}07^{\mathrm{s}}\!.99$ $+40^{\circ}55^{\prime}01^{\prime\prime}\!\!.1$ (J2000) with an unfiltered magnitude of 19.0 \citep{2015ATel.6911....1S}. The transient then brightened significantly from $R=18.00\pm0.03$ on January 13.74 to $R=16.82\pm0.02$ on January 14.71 \citep{2015ATel.6924....1O}. A spectrum was taken by \citet{2015ATel.6924....1O} on 2015 January 14.8, which revealed the presence of H$\alpha$ emission, leading to the tentative classification of the transient as a classical nova in M31.  Additional observations and analyses have been reported by \citet{2015ATel.7173....1D}, \citet{2015ATel.7208....1W}, and \citet{2015Ada}.

As part of our extensive optical follow-up campaign for M31 novae (e.g.\ \citealp{2011ApJ...734...12S,2014ApJS..213...10W}), we observed M31LRN~2015 photometrically and spectroscopically with the Liverpool Telescope (LT), a fully robotic 2m telescope on the island of La Palma.  

\section{Photometric Observations}\label{photobs}
\begin{figure}[ht]
\centering\includegraphics[width=0.9\columnwidth]{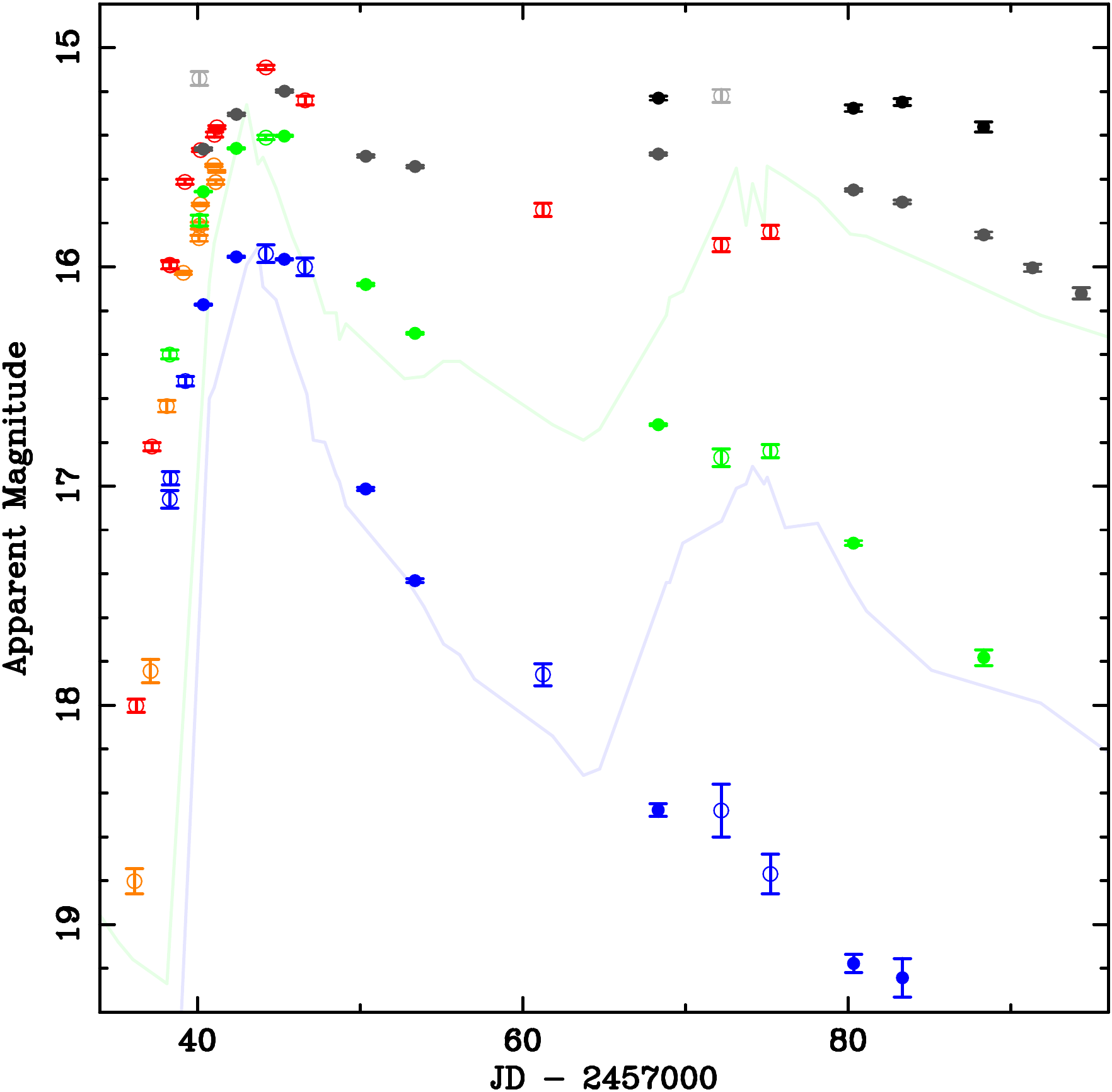}
\caption{Light curve of M31LRN~2015 from LT observations (filled circles), as well as data from \citet{2015ATel.6924....1O}, \citet{2015ATel.6941....1K,2015ATel.7150....1K}, \citet{2015ATel.6951....1S} and \citet{2015ATel.6985....1F} (all open circles). {\it B}-band data is represented by blue points; {\it V}, green; {\it W}, orange; {\it R}, red; {\it I}, light gray; $i'$, dark gray; $z'$, black. The light blue and green lines represent the {\it B} and {\it V} photometry respectively of the V838 Mon outburst from \citet{2002A&A...389L..51M} shifted to the distance of M31 using a distance to V838 Mon of 6.1~kpc \citep{2008AJ....135..605S} and the extinction of $E_{(B-V)}=0.7$ (see \citealp{2005MNRAS.360.1281R}).\label{lc}}
\end{figure}

We obtained multi-color photometric observations of M31LRN~2015 using the IO:O CCD camera on the LT \citep{2004SPIE.5489..679S}. Typically $3\times120$~s exposures were taken in each filter, which were decreased to $3\times60$~s at later times as M31 became more difficult to observe. The {\it B}, {\it V} and $i'$-band photometric data were calibrated using eight stars from \citet{2006AJ....131.2478M}, with the $i'$-band magnitudes computed using transformations from \citet{2006A&A...460..339J}. The $z'$-band data were calibrated using a single SDSS DR9 \citep{2012ApJS..203...21A} star (the only star in the field suitable for the calibration), J004225.86+405651.4, which may have affected the reliability, but by comparison to the $i'$-band photometry, we determined any systematic effect is small ($<0.1$~mag). The photometry is presented in Table~\ref{phot} and a light curve combining published data with those taken with the LT is shown in Figure~\ref{lc}.

\begin{deluxetable*}{lllll}
\tablecaption{LT photometry of M31LRN 2015.\label{phot}}
\tablewidth{0pt}
\tablehead{
\colhead{Date (UT)} & \colhead{{\it B}} & \colhead{{\it V}} & \colhead{$i'$} & \colhead{$z'$}
}
\startdata
2015 Jan 17.87 &$16.172\pm0.003$ &$15.657\pm0.002$ &$15.463\pm0.006$ &\nodata\\
2015 Jan 19.89 &$15.954\pm0.004$ &$15.459\pm0.003$ &$15.304\pm0.006$ &\nodata\\
2015 Jan 22.85 &$15.965\pm0.003$ &$15.403\pm0.002$ &$15.198\pm0.005$ &\nodata\\
2015 Jan 27.85 &$17.013\pm0.008$ &$16.080\pm0.005$ &$15.495\pm0.006$ &\nodata\\
2015 Jan 30.88 &$17.431\pm0.009$ &$16.303\pm0.004$ &$15.542\pm0.006$ &\nodata\\
2015 Feb 14.83 &$18.478\pm0.029$ &$16.720\pm0.004$ &$15.485\pm0.006$ &$15.230\pm0.009$\\
2015 Feb 26.84 &$19.178\pm0.041$ &$17.260\pm0.011$ &$15.649\pm0.006$ &$15.276\pm0.015$\\
2015 Mar 1.83 &$19.243\pm0.088$ &\nodata &$15.704\pm0.009$ &$15.248\pm0.015$\\
2015 Mar 6.83 &\nodata &$17.782\pm0.036$ &$15.854\pm0.014$ &$15.361\pm0.023$\\
2015 Mar 9.84 &\nodata &\nodata &$16.004\pm0.017$ &\nodata\\
2015 Mar 12.83 &\nodata &\nodata &$16.121\pm0.025$ &\nodata
\enddata
\tablecomments{{\it BV$i'$} calibrated against stars from \citet{2006AJ....131.2478M}, $z'$ calibrated against star from SDSS.}
\end{deluxetable*}

The most striking feature of the light curve is the increasingly red color after peak, from ($B-i'$) = $0.650\pm0.007$ on January 19.89 to ($B-i'$) = $3.54\pm0.09$ on March 1.83. Our data imply optical peak was reached between 2015 January 19.89 and January 22.84. M31 has a distance modulus of ({\it m$-$M})$_0 = 24.43\pm0.06$ \citep{1990ApJ...365..186F}. If we assume foreground reddening of $E_{(B-V)}=0.062$ \citep{1998ApJ...500..525S} and maximum total extinction at that position in M31 of $E_{(B-V)}=0.18$ \citep{2009A&A...507..283M}, we conservatively assume that M31LRN~2015 was subject to reddening of $E_{(B-V)}=0.12\pm0.06$. As such, the absolute peak magnitude of the transient was $M_V=-9.4\pm0.2$. It is possible that the peak magnitude was slightly brighter, although probably not significantly, given that the available photometric data in all filters seems to suggest a smooth transition from the rise to decline phases of the light curve.

\section{Spectroscopic Observations} \label{sec:spec}
We obtained four spectra of M31LRN~2015 using the SPRAT spectrograph on the LT in red optimized mode, with a resolution of 18\,\AA, under photometric conditions. A spectrophotometric standard (Hiltner 600; \citealp{1977ApJ...218..767S}) was observed on the night of 2015 February 4 in $1^{\prime\prime}\!\!.4$ seeing (FWHM) and used to flux calibrate the data. SPRAT has a fixed width slit equivalent to $1^{\prime\prime}\!\!.8$ on the sky. Seeing in the science spectra varies in the range $1^{\prime\prime}\!\!.5-2^{\prime\prime}\!\!.5$ (FWHM) between nights. This seeing variation forms the major cause of uncertainty in our absolute flux calibration, which we estimate as $\sim25\%$. However since all observations were obtained with the slit aligned with the parallactic angle, the relative flux calibration within a single spectrum is better constrained (typically $<10\%$ between 5000 and 7500\,\AA\ for all data).

\begin{figure*}[ht]
\centering\includegraphics[width=0.8\textwidth]{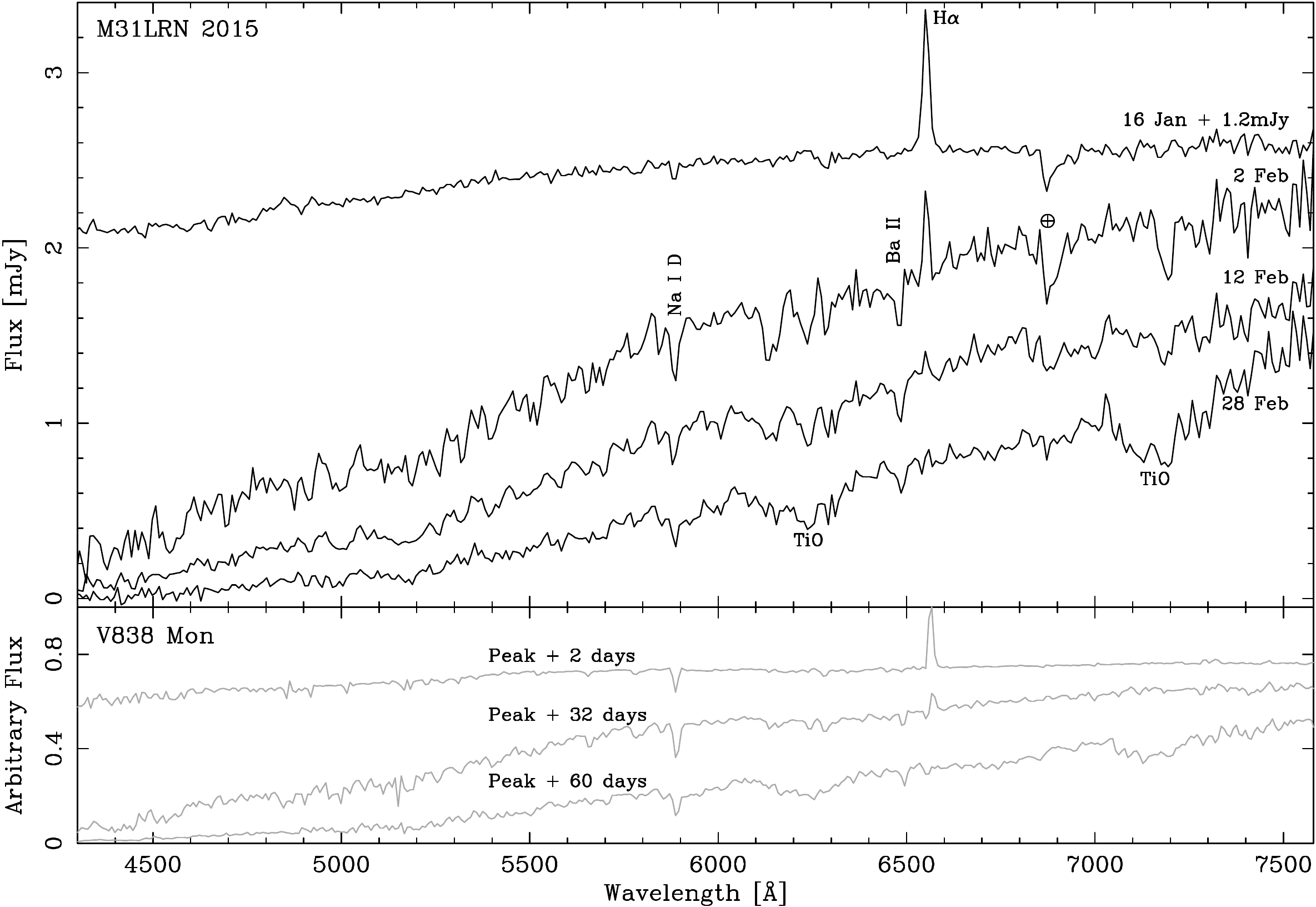}
\caption{Top: the four LT SPRAT spectra of M31LRN~2015 taken 2015 January 16.86, February 2.86, February 12.84 and February 28.85, with peak occurring between 2015 January 19.89 and January 22.84 (see text). Bottom: three spectra of V838 Mon from \citet{2005MNRAS.360.1281R} with respect to its peak, binned to a similar resolution as our SPRAT spectra of M31LRN~2015.\label{spec}}
\end{figure*}

The first spectrum was obtained on 2015 January 16.86, 3.2~days after discovery, and prior to peak. The only strong feature is the H$\alpha$ emission,  which has a FWHM of $900\pm100$\,km\,s$^{-1}$ and peaks at $6551.8\pm0.5$\,\AA, consistent with an M31 origin. We derive a H$\alpha$ flux of $(7.8\pm0.7)\times10^{-11}$\,erg\,cm$^{-2}$\,s$^{-1}$ from the best-fit Gaussian profile. A small amount of Na~{\sc i} D absorption is detected (FWHM = $600\pm200$\,km\,s$^{-1}$), although this could be interstellar in origin, as well as H$\beta$ being tentatively identified. This, along with the other SPRAT spectra taken of M31LRN~2015, is shown in Figure~\ref{spec}. 

The second spectrum was obtained 17~days later on February 2.86. In this spectrum, the H$\alpha$ flux had weakened to $(3.6\pm1.0)\times10^{-11}$\,erg\,cm$^{-2}$\,s$^{-1}$, the continuum is much redder, Na~{\sc i} D absorption is now strong and several other strong absorption features have emerged. The third spectrum, taken February 12.84, is somewhat similar, but H$\alpha$ is now only just visible above the continuum (which is even redder). The most prominent absorption features in the second and third spectra are at $5882\pm1$\,\AA\ (Na~{\sc i} D; FWHM = $1000\pm200$\,km\,s$^{-1}$), $6131\pm3$, $6240\pm1$, and $6482\pm1$\,\AA. The $6482\pm1$\,\AA\ absorption line (FWHM = $900\pm200$\,km\,s$^{-1}$) is likely to be Ba~{\sc ii} (6496.9\,\AA) blueshifted by $690\pm50$\,km\,s$^{-1}$, making the more poorly constrained $6131\pm3$\,\AA\ line consistent with Ba~{\sc ii} (6141.7\,\AA). There is also a less prominent absorption feature blue-ward of Na~{\sc i} D with a minimum at 5841.6\,\AA, but a profile could not be fit, which is possibly caused by Ba~{\sc ii} (5853.7\,\AA). The absorption at $6240\pm1$\,\AA, may be caused by Fe~{\sc ii}, as seen (along with Ba~{\sc ii}) in the V838 Mon spectra published by \citet{2002AstL...28..691G}. We estimate the Na~{\sc i} D equivalent width (EW) as $1.2\pm0.3$\,\AA\ on January 16.86 and $5\pm1$\,\AA\ on February 12.84. Using the relation from \citet{2011MNRAS.415L..81P}, this gives a rough approximation of  $E_{(B-V)} \sim 0.4$ on January 16.86 and  $E_{(B-V)} \sim 2$ on February 12.84. Note however that those authors warn against using Na~{\sc i} D EW in low-resolution spectra to estimate absolute extinction, but it does at least show that the absorbing column may be increasing between the two epochs.

The fourth spectrum was taken February 28.85 and showed significant changes, with the spectrum now somewhat resembling that of a K or M-type supergiant. No significant H$\alpha$ emission is detected and there are clear TiO absorption bands, with the Ba~{\sc ii} and Na~{\sc i} D lines still present. A spectrum was also taken on February 25.84 but is not included as it was very similar to the one taken February 28, but noisier.

\section{The Progenitor System} \label{sec:prog}
Having successfully recovered a number of progenitor systems for M31 novae \citep{2009ApJ...705.1056B,2014A&A...563L...9D,2014ApJS..213...10W}, we used the same method to search for a possible progenitor system of M31LRN~2015. The archival {\it Hubble Space Telescope (HST)} observations were taken using the Advanced Camera for Surveys Wide Field Channel (ACS/WFC) through F555W and F814W filters on 2004 August 16.5 (prop.\ ID: 10273). Three $i'$-band LT frames taken on 2015 January 17.87 were stacked and 37 bright stars common to both observations were used to compute the geometric transformation between the two datasets using IRAF \citep{1986SPIE..627..733T} routines. This transformation was used to determine the position of the quiescent M31LRN~2015 in the archival {\it HST} data.

\begin{figure}[ht]
\centering\includegraphics[width=0.9\columnwidth]{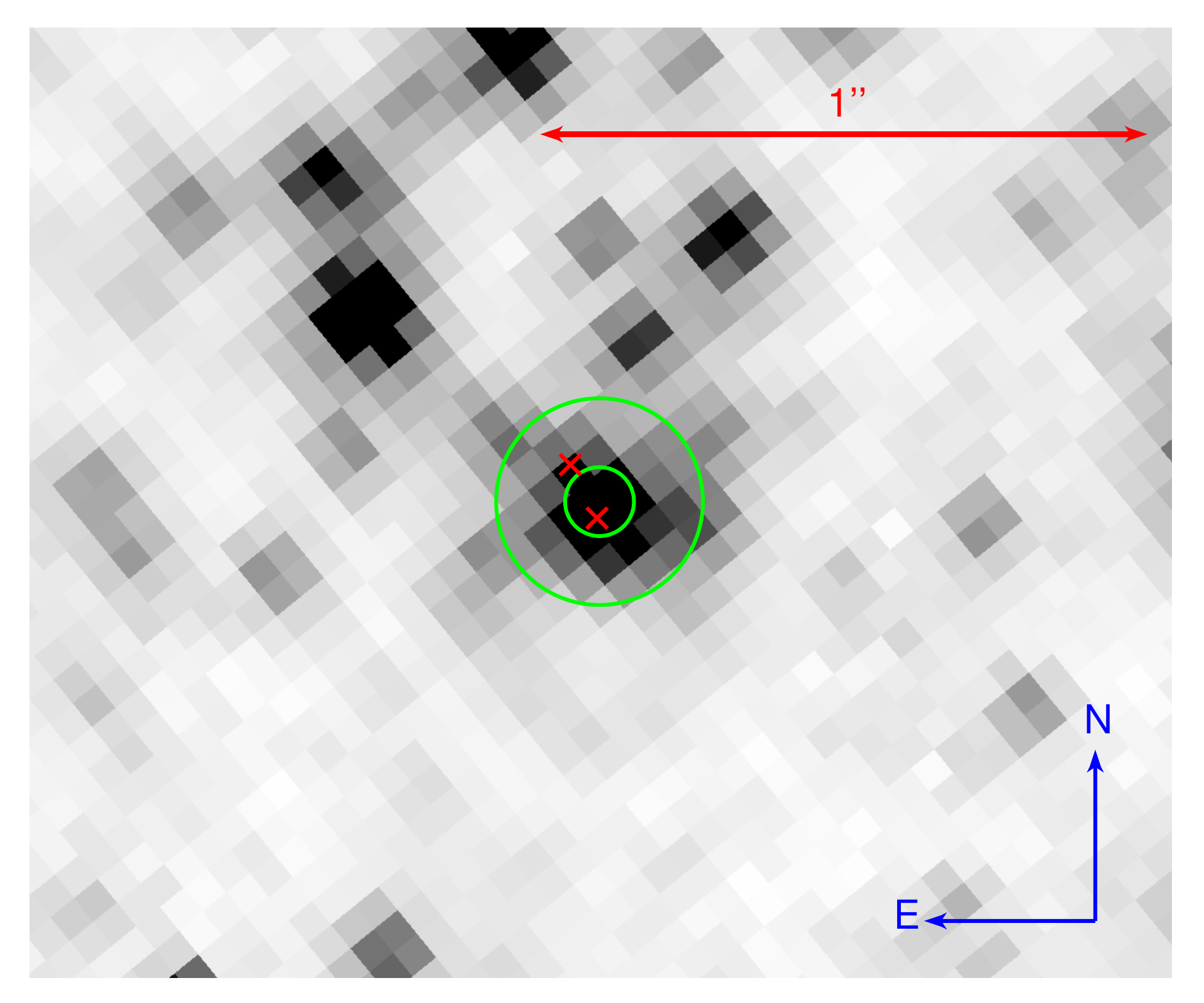}
\caption{F814W {\it HST} (negative) image of the location of M31LRN~2015 taken 2004 August 16.5. The 1$\sigma$ and 3$\sigma$ errors on the calculated position of the quiescent system are represented by green circles. The red crosses represent the positions of the nearby sources in the {\it HST} data.\label{prog}}
\end{figure}

This analysis reveals a source within 1$\sigma$ of the position of M31LRN~2015 in the {\it HST} data. Photometry was performed on all sources using DOLPHOT \citep[v2.0\footnote{\url{http://purcell.as.arizona.edu/dolphot}};][]{2000PASP..112.1383D}. The closest source to the calculated position of M31LRN~2015, the progenitor candidate, is located at a distance of 0.55 ACS/WFC pixels, 0.5$\sigma$ or $0^{\prime\prime}\!\!.027$, and has magnitudes of F555W $=23.43\pm0.04$ and F814W $=22.09\pm0.01$ \citep[these are in agreement with a similar analysis by][]{2015ATel.7173....1D}. We also note there is an additional source 1.4$\sigma$ or $0^{\prime\prime}\!\!.077$ away. This second source has a magnitude of F814W $=24.48\pm0.07$, but is undetected in the F555W data. The calculated position of M31LRN~2015, along with those nearby sources are shown in Figure~\ref{prog}. Using a similar method to that described in \citet{2014ApJS..213...10W}, we determine the probability of a source being within $0^{\prime\prime}\!\!.027$ by chance is just $3\%$, given the resolved local stellar density.

Employing the conversions from \citet{2005PASP..117.1049S} and using the distance and reddening toward M31LRN~2015 in M31 (see Section~\ref{photobs}) we estimate the absolute magnitude of the progenitor candidate to be $M_V=-1.50\pm0.23$, $M_I=-2.55\pm0.13$ with ($V-I$)$_{0}=1.05\pm0.15$, consistent with a red giant.

The Local Group Galaxies Survey (LGGS; \citealp{2006AJ....131.2478M}) data reveal a nearby source, J004208.06+405501.7, at $V=22.33\pm0.09$, ($B-V$) $=0.75\pm0.17$, ($V-R$) $=0.62\pm0.10$ and ($R-I$) $=0.65\pm0.04$. While it would first appear that the progenitor is brighter in the \citet{2006AJ....131.2478M} catalog, an inspection of the images reveals this source is in fact a blend of several of the stars visible in Figure~\ref{prog}, including the two brightest. A source is clearly visible at the position of M31LRN~2015 in the LGGS narrowband H$\alpha$ data (taken 2002 September 11), but nothing is seen at this position in either the [O~{\sc iii}] or [S~{\sc ii}] data. To remove any contribution from continuum emission in these narrowband images we subtracted the scaled LGGS {\it R}-band image from both the H$\alpha$ and [S~{\sc ii}] images, and the scaled {\it V}-band image from the [O~{\sc iii}] image. The scaling was calculated by determining a linear fit between the broadband and narrowband data using SExtractor (v2.19.5; \citealp{1996A&AS..117..393B}) photometry of $\sim20,000$ objects in each image. A clear source was seen in the continuum-subtracted H$\alpha$ data, indicating a strong H$\alpha$ excess in the spectrum of the pre-outburst M31LRN~2015. \citet{2011AJ....142..139A}, who also used the LGGS data, list it as an H~{\sc ii} region (\#1527 in their paper), with an estimated luminosity of $6\times10^{34}$\,erg\,s$^{-1}$. If indeed this is a star forming H~{\sc ii} region, the luminosity implies a star formation rate of $\sim5\times10^{-7}$~M$_{\odot}$\,yr$^{-1}$ (using the conversion from \citealp{1998ARA&A..36..189K}). A SN remnant origin for this emission seems unlikely, due to the lack of [S~{\sc ii}] emission and the small size, which implies an age of less than a few hundred years. If this is not an H~{\sc ii} region, the most likely source of the H$\alpha$ emission is the progenitor system itself, either directly from the progenitor or a companion, or from a period of mass loss.

\section{Discussion}
M31LRN~2015 is a luminous, red transient in M31, characterized by weakening H$\alpha$ emission on an increasingly red continuum, with a number of absorption lines emerging after peak, including Na~{\sc i} D and Ba~{\sc ii}, and TiO bands appearing at later times. Considering the spectroscopic and photometric evolution of the outburst we conclude that this object is a LRN in M31, a conclusion that \citet{2015ATel.7150....1K} reached independently from their data. We now compare the properties of this system to those of other proposed LRNe and ILRTs.

The light curves of both LRNe and ILRTs are characterized by slow evolution (compared to most novae) and significant reddening as they fade. The light curves of LRNe V1309 Sco \citep{2010A&A...516A.108M} and V838 Mon \citep{2002A&A...389L..51M} show the brightness quickly falling in the bluer filters, but plateauing in the red filters. The light curve of V4332 Sagittarii (another LRN; \citealp{1999AJ....118.1034M}) also shows this object to be reddening about two weeks after discovery, although there is no color information for the earlier periods. The only LRN that shows clear evidence of a significant multiple-peak light curve structure is V838 Mon. ILRTs are more luminous at peak than the LRNe discussed above and they also show different photometric evolution. The first ILRT discovered, M85~OT2006-1, shows a long plateau phase ($\sim70$~days), with relatively little color evolution until later times \citep{2007Natur.447..458K}. ILRT PTF~10fqs showed a plateau phase ($\sim40$~days), before rapidly reddening \citep{2011ApJ...730..134K}, and NGC 300 OT2008 also shows no major optical color evolution until later times \citep{2009ApJ...695L.154B}. Our observations show that M31LRN~2015 clearly resembles LRNe, rather than ILRTs, both in peak luminosity and light curve evolution. In the late-time M31LRN~2015 light curve (see Figure~\ref{lc}), the $i'$-band plateau appears to be ending, giving a plateau time of 40$-$50~days. Comparing the plateau time and luminosity to the common-envelope event theoretical models of \citet[][see their Figure 1]{2013Sci...339..433I}, M31LRN~2015 is consistent with the model for a merger with ejecta mass of 0.1~M$_{\odot}$, but is also consistent with an event where the entire envelope is ejected.

The spectra of LRNe and ILRTs do show similarities, with both types of object exhibiting strong Na~{\sc i}~D absorption and narrow H$\alpha$ emission, but they also appear to show some differences. Strong [Ca~{\sc ii}] emission has been seen in several ILRTs (see e.g.\ \citealp{2011ApJ...730..134K}), which is not the case for LRNe. Additionally, strong TiO absorption bands were seen at later times in LRNe V4332 Sgr \citep{1999AJ....118.1034M} and V838 Mon (see e.g.\ \citealp{2005MNRAS.360.1281R}). These were also seen in M31-RV \citep{1989ApJ...341L..51R}. V838 Mon is the LRN with the best spectroscopic coverage, and the evolution of M31LRN~2015 seems to match that very well, for example,  Ba~{\sc ii} (6496.9\,\AA) appears between $\sim2$~days prior to peak ($t\sim-2$~days; on 2002 February 5) and $t\sim+9$~days in V838 Mon \citep{2002AstL...28..691G}, and also appears in M31LRN~2015 after peak. It can be seen from Figure~\ref{spec} that our second and third spectra of M31LRN~2015 are remarkably similar to spectra taken around $t\sim+32$~days of V838 Mon published by \citet{2005MNRAS.360.1281R}, with our fourth spectrum of M31LRN~2015 similar to V838 Mon at $t\sim+60$~days.

The only significant differences between M31LRN~2015 and V838 Mon are the pre-outburst detections of the progenitor systems and the second bright peak in the light curve of the latter. The V838 Mon system contains a young main-sequence binary companion \citep{2002IAUC.8005....2M}. For the case of M31LRN~2015, we may again be seeing a binary companion, but at a more evolved stage, and depending on the mechanism, this may not represent different progenitor stars for the outbursts themselves. Notably, no young stellar population similar to the locale of V838 Mon was found to be coincident with M31-RV \citep{2006AJ....131..984B}. If the giant planet capture mechanism was responsible for LRNe outbursts, this system (along with M31-RV) would represent the first (albeit indirect) detection of extragalactic planetary systems. While the detected progenitor candidate does not exclude a classical nova mechanism \citep{2012ApJ...746...61D,2014ApJS..213...10W} for M31LRN~2015, the presence of a high-mass main-sequence star in the V838 Mon system and the strong evidence for a stellar merger in the case of V1309 Sco appears to rule it out for those systems. The spectra are not consistent with a typical nova eruption, although \citet{2010ApJ...725..831S} offer an explanation for this. The strong evidence for the V1309 Sco outburst being caused by the merger of a contact binary \citep{2011A&A...528A.114T} suggests that if a single progenitor mechanism is responsible for all LRNe, this is likely to be the stellar merger scenario. \citet{2015ATel.7173....1D} reported possible pre-outburst variability in M31LRN 2015 from difference image analysis (DIA) of archival data.  Although the presence of many faint sources in such crowded fields (see Section~\ref{sec:prog}) can affect the reliability of photometry, even when employing DIA, it is unlikely in this case and the significant brightening they detect from $g\sim22.6$ to $g\sim20.8$ between 2009 and 2014 certainly warrants further investigation.

\section{Conclusions}
In this Letter we have presented photometric and spectroscopic observations of M31LRN~2015, and a search for the progenitor system in broad and narrowband archival data. Here we summarize the main conclusions.
\begin{enumerate}
\item We used the LT to photometrically and spectroscopically observe M31LRN~2015, which is a LRN peaking at an absolute magnitude of $M_V=-9.4\pm0.2$. It took several days to reach optical peak, with the optical colors becoming increasingly redder over the following weeks.
\item The spectra are characterized by a reddening continuum, weakening H$\alpha$ emission and emerging absorption features, including Na~{\sc i} D and Ba~{\sc ii}. At later times TiO absorption bands are also seen. M31LRN~2015 shows a number of similarities with the spectra of other transients classified as LRNe, resembling V838 Mon particularly closely.
\item The likely progenitor system has an absolute magnitude of $M_V=-1.50\pm0.23$ with ($V-I$)$_{0}=1.05\pm0.15$, although this could represent photometry of a more luminous companion star. Pre-outburst H$\alpha$ emission is likely to be related to the progenitor system, either directly or the local environment.
\item The nature of all of the LRN class of objects is still uncertain, but our observations of M31LRN~2015 and its likely progenitor system suggest that the mechanism(s) must produce very similar outbursts (in peak luminosity, timescale and outburst evolution), and possibly within differing stellar populations. If a single progenitor mechanism is responsible for all LRNe, the balance of evidence favors the stellar merger pathway.
\end{enumerate}

Further observations, particularly of the post-outburst system, may yield further clues to the nature of M31LRN~2015 and the LRN class of objects in general.

\acknowledgements We would like to thank M.~Rushton for supplying the V838 Mon spectra used here for comparison with M31LRN~2015. We also thank an anonymous referee for the constructive suggestions. The Liverpool Telescope is operated on the island of La Palma by Liverpool John Moores University in the Spanish Observatorio del Roque de los Muchachos of the Instituto de Astrofisica de Canarias with financial support from the UK Science and Technology Facilities Council.  Some of the data presented in this paper were obtained from the Multimission Archive at the Space Telescope Science Institute (MAST).  STScI is operated by the Association of Universities for Research in Astronomy, Inc., under NASA contract NAS5-26555.  Support for MAST for non-{\it HST} data is provided by the NASA Office of Space Science via grant NNX09AF08G and by other grants and contracts.

{\it Facilities}: \facility{{\it HST}}, \facility{Liverpool:2m}.
\bibliographystyle{apj}

\begin{thebibliography}{}
\expandafter\ifx\csname natexlab\endcsname\relax\def\natexlab#1{#1}\fi

\scriptsize {

\bibitem[{{Adams} {et~al.}(2015){Adams}, {Kochanek}, {Dong}, \& {Wagner}}]{2015Ada}
{Adams}, S., {Kochanek}, C.~S., {Dong}, S., \& {Wagner},
  R.~M. 2015, ATel, 7468

\bibitem[{{Ahn} {et~al.}(2012){Ahn}, {Alexandroff}, {Allende Prieto},
  {Anderson}, {Anderton}, {Andrews}, {Aubourg}, {Bailey}, {Balbinot}, {Barnes},
  \& et~al.}]{2012ApJS..203...21A}
{Ahn}, C.~P., {Alexandroff}, R., {Allende Prieto}, C., {et~al.} 2012, \apjs,
  203, 21

\bibitem[{{Azimlu} {et~al.}(2011){Azimlu}, {Marciniak}, \&
  {Barmby}}]{2011AJ....142..139A}
{Azimlu}, M., {Marciniak}, R., \& {Barmby}, P. 2011, \aj, 142, 139

\bibitem[{{Bertin} \& {Arnouts}(1996)}]{1996A&AS..117..393B}
{Bertin}, E., \& {Arnouts}, S. 1996, \aaps, 117, 393

\bibitem[{{Bode} {et~al.}(2009){Bode}, {Darnley}, {Shafter}, {Page},
  {Smirnova}, {Anupama}, \& {Hilton}}]{2009ApJ...705.1056B}
{Bode}, M.~F., {Darnley}, M.~J., {Shafter}, A.~W., {et~al.} 2009, \apj, 705,
  1056

\bibitem[{{Bond} {et~al.}(2009){Bond}, {Bedin}, {Bonanos}, {Humphreys},
  {Monard}, {Prieto}, \& {Walter}}]{2009ApJ...695L.154B}
{Bond}, H.~E., {Bedin}, L.~R., {Bonanos}, A.~Z., {et~al.} 2009, \apjl, 695,
  L154

\bibitem[{{Bond} \& {Siegel}(2006)}]{2006AJ....131..984B}
{Bond}, H.~E., \& {Siegel}, M.~H. 2006, \aj, 131, 984

\bibitem[{{Bond} {et~al.}(2003){Bond}, {Henden}, {Levay}, {Panagia}, {Sparks},
  {Starrfield}, {Wagner}, {Corradi}, \& {Munari}}]{2003Natur.422..405B}
{Bond}, H.~E., {Henden}, A., {Levay}, Z.~G., {et~al.} 2003, \nat, 422, 405

\bibitem[{{Darnley} {et~al.}(2012){Darnley}, {Ribeiro}, {Bode}, {Hounsell}, \&
  {Williams}}]{2012ApJ...746...61D}
{Darnley}, M.~J., {Ribeiro}, V.~A.~R.~M., {Bode}, M.~F., {Hounsell}, R.~A., \&
  {Williams}, R.~P. 2012, \apj, 746, 61

\bibitem[{{Darnley} {et~al.}(2014){Darnley}, {Williams}, {Bode}, {Henze},
  {Ness}, {Shafter}, {Hornoch}, \& {Votruba}}]{2014A&A...563L...9D}
{Darnley}, M.~J., {Williams}, S.~C., {Bode}, M.~F., {et~al.} 2014, \aap, 563,
  L9

\bibitem[{{Dolphin}(2000)}]{2000PASP..112.1383D}
{Dolphin}, A.~E. 2000, \pasp, 112, 1383

\bibitem[{{Dong} {et~al.}(2015){Dong}, {Kochanek}, {Adams}, \&
  {Prieto}}]{2015ATel.7173....1D}
{Dong}, S., {Kochanek}, C.~S., {Adams}, S., \& {Prieto}, J.-L. 2015, ATel, 7173

\bibitem[{{Fabrika} {et~al.}(2015){Fabrika}, {Barsukova}, {Valeev},
  {Vinokurov}, {Sholukhova}, {Goranskij}, {Hornoch}, {Henze}, \&
  {Shafter}}]{2015ATel.6985....1F}
{Fabrika}, S., {Barsukova}, E.~A., {Valeev}, A.~F., {et~al.} 2015, ATel, 6985

\bibitem[{{Freedman} \& {Madore}(1990)}]{1990ApJ...365..186F}
{Freedman}, W.~L., \& {Madore}, B.~F. 1990, \apj, 365, 186

\bibitem[{{Goranskii} {et~al.}(2002){Goranskii}, {Kusakin}, {Metlova},
  {Shugarov}, {Barsukova}, \& {Borisov}}]{2002AstL...28..691G}
{Goranskii}, V.~P., {Kusakin}, A.~V., {Metlova}, N.~V., {et~al.} 2002,
  Astronomy Letters, 28, 691

\bibitem[{{Ivanova} {et~al.}(2013){Ivanova}, {Justham}, {Avendano Nandez}, \&
  {Lombardi}}]{2013Sci...339..433I}
{Ivanova}, N., {Justham}, S., {Avendano Nandez}, J.~L., \& {Lombardi}, J.~C.
  2013, Science, 339, 433

\bibitem[{{Jordi} {et~al.}(2006){Jordi}, {Grebel}, \&
  {Ammon}}]{2006A&A...460..339J}
{Jordi}, K., {Grebel}, E.~K., \& {Ammon}, K. 2006, \aap, 460, 339

\bibitem[{{Kasliwal} {et~al.}(2011){Kasliwal}, {Kulkarni}, {Arcavi}, {Quimby},
  {Ofek}, {Nugent}, {Jacobsen}, {Gal-Yam}, {Green}, {Yaron}, {Fox}, {Howell},
  {Cenko}, {Kleiser}, {Bloom}, {Miller}, {Li}, {Filippenko}, {Starr},
  {Poznanski}, {Law}, {Helou}, {Frail}, {Neill}, {Forster}, {Martin},
  {Tendulkar}, {Gehrels}, {Kennea}, {Sullivan}, {Bildsten}, {Dekany}, {Rahmer},
  {Hale}, {Smith}, {Zolkower}, {Velur}, {Walters}, {Henning}, {Bui}, {McKenna},
  \& {Blake}}]{2011ApJ...730..134K}
{Kasliwal}, M.~M., {Kulkarni}, S.~R., {Arcavi}, I., {et~al.} 2011, \apj, 730,
  134

\bibitem[{{Kennicutt}(1998)}]{1998ARA&A..36..189K}
{Kennicutt}, Jr., R.~C. 1998, \araa, 36, 189

\bibitem[{{Kochanek}(2011)}]{2011ApJ...741...37K}
{Kochanek}, C.~S. 2011, \apj, 741, 37

\bibitem[{{Kulkarni} {et~al.}(2007){Kulkarni}, {Ofek}, {Rau}, {Cenko},
  {Soderberg}, {Fox}, {Gal-Yam}, {Capak}, {Moon}, {Li}, {Filippenko}, {Egami},
  {Kartaltepe}, \& {Sanders}}]{2007Natur.447..458K}
{Kulkarni}, S.~R., {Ofek}, E.~O., {Rau}, A., {et~al.} 2007, \nat, 447, 458

\bibitem[{{Kurtenkov} {et~al.}(2015{\natexlab{a}}){Kurtenkov}, {Ovcharov},
  {Nedialkov}, {Kostov}, {Bachev}, {Dimitrova}, {Popov}, \&
  {Valcheva}}]{2015ATel.6941....1K}
{Kurtenkov}, A., {Ovcharov}, E., {Nedialkov}, P., {et~al.} 2015{\natexlab{a}},
  ATel, 6941

\bibitem[{{Kurtenkov} {et~al.}(2015{\natexlab{b}}){Kurtenkov}, {Tomov},
  {Fabrika}, {Valeev}, {Pessev}, {Vida}, \& {Molnar}}]{2015ATel.7150....1K}
{Kurtenkov}, A., {Tomov}, T., {Fabrika}, E.~A., {et~al.} 2015{\natexlab{b}},
  ATel, 7150

\bibitem[{{Lovegrove} \& {Woosley}(2013)}]{2013ApJ...769..109L}
{Lovegrove}, E., \& {Woosley}, S.~E. 2013, \apj, 769, 109

\bibitem[{{Martini} {et~al.}(1999){Martini}, {Wagner}, {Tomaney}, {Rich},
  {della Valle}, \& {Hauschildt}}]{1999AJ....118.1034M}
{Martini}, P., {Wagner}, R.~M., {Tomaney}, A., {et~al.} 1999, \aj, 118, 1034

\bibitem[{{Mason} {et~al.}(2010){Mason}, {Diaz}, {Williams}, {Preston}, \&
  {Bensby}}]{2010A&A...516A.108M}
{Mason}, E., {Diaz}, M., {Williams}, R.~E., {Preston}, G., \& {Bensby}, T.
  2010, \aap, 516, A108

\bibitem[{{Massey} {et~al.}(2006){Massey}, {Olsen}, {Hodge}, {Strong},
  {Jacoby}, {Schlingman}, \& {Smith}}]{2006AJ....131.2478M}
{Massey}, P., {Olsen}, K.~A.~G., {Hodge}, P.~W., {et~al.} 2006, \aj, 131, 2478

\bibitem[{{Montalto} {et~al.}(2009){Montalto}, {Seitz}, {Riffeser}, {Hopp},
  {Lee}, \& {Sch{\"o}nrich}}]{2009A&A...507..283M}
{Montalto}, M., {Seitz}, S., {Riffeser}, A., {et~al.} 2009, \aap, 507, 283

\bibitem[{{Munari} {et~al.}(2002{\natexlab{a}}){Munari}, {Desidera}, \&
  {Henden}}]{2002IAUC.8005....2M}
{Munari}, U., {Desidera}, S., \& {Henden}, A. 2002{\natexlab{a}}, \iaucirc,
  8005, 2

\bibitem[{{Munari} {et~al.}(2002{\natexlab{b}}){Munari}, {Henden}, {Kiyota},
  {Laney}, {Marang}, {Zwitter}, {Corradi}, {Desidera}, {Marrese}, {Giro},
  {Boschi}, \& {Schwartz}}]{2002A&A...389L..51M}
{Munari}, U., {Henden}, A., {Kiyota}, S., {et~al.} 2002{\natexlab{b}}, \aap,
  389, L51

\bibitem[{{Ovcharov} {et~al.}(2015){Ovcharov}, {Kurtenkov}, {Valcheva}, \&
  {Nedialkov}}]{2015ATel.6924....1O}
{Ovcharov}, E., {Kurtenkov}, A., {Valcheva}, A., \& {Nedialkov}, P. 2015, ATel, 6924

\bibitem[{{Poznanski} {et~al.}(2011){Poznanski}, {Ganeshalingam}, {Silverman},
  \& {Filippenko}}]{2011MNRAS.415L..81P}
{Poznanski}, D., {Ganeshalingam}, M., {Silverman}, J.~M., \& {Filippenko},
  A.~V. 2011, \mnras, 415, L81

\bibitem[{{Retter} \& {Marom}(2003)}]{2003MNRAS.345L..25R}
{Retter}, A., \& {Marom}, A. 2003, \mnras, 345, L25

\bibitem[{{Rich} {et~al.}(1989){Rich}, {Mould}, {Picard}, {Frogel}, \&
  {Davies}}]{1989ApJ...341L..51R}
{Rich}, R.~M., {Mould}, J., {Picard}, A., {Frogel}, J.~A., \& {Davies}, R.
  1989, \apjl, 341, L51

\bibitem[{{Rushton} {et~al.}(2005){Rushton}, {Geballe}, {Filippenko},
  {Chornock}, {Li}, {Leonard}, {Foley}, {Evans}, {Smalley}, {van Loon}, \&
  {Eyres}}]{2005MNRAS.360.1281R}
{Rushton}, M.~T., {Geballe}, T.~R., {Filippenko}, A.~V., {et~al.} 2005, \mnras,
  360, 1281

\bibitem[{{Schlegel} {et~al.}(1998){Schlegel}, {Finkbeiner}, \&
  {Davis}}]{1998ApJ...500..525S}
{Schlegel}, D.~J., {Finkbeiner}, D.~P., \& {Davis}, M. 1998, \apj, 500, 525

\bibitem[{{Shafter} {et~al.}(2011){Shafter}, {Darnley}, {Hornoch},
  {Filippenko}, {Bode}, {Ciardullo}, {Misselt}, {Hounsell}, {Chornock}, \&
  {Matheson}}]{2011ApJ...734...12S}
{Shafter}, A.~W., {Darnley}, M.~J., {Hornoch}, K., {et~al.} 2011, \apj, 734, 12

\bibitem[{{Shara} {et~al.}(2010){Shara}, {Yaron}, {Prialnik}, {Kovetz}, \&
  {Zurek}}]{2010ApJ...725..831S}
{Shara}, M.~M., {Yaron}, O., {Prialnik}, D., {Kovetz}, A., \& {Zurek}, D. 2010,
  \apj, 725, 831

\bibitem[{{Shumkov} {et~al.}(2015{\natexlab{a}}){Shumkov}, {Pruzhinskaya},
  {Tiurina}, {Lipunov}, {Gorbovskoy}, {Balanutsa}, {Denisenko}, {Vladimirov},
  {Rufanov}, {Kuznetsov}, {Buckley}, {Potter}, {Kniazev}, {Kotze}, {Ivanov},
  {Gres}, {Yazev}, {Budnev}, {Poleshchuk}, {Konstantinov}, {Tlatov},
  {Dormidontov}, {Senik}, {Parkhomenko}, {Yurkov}, {Sergienko}, {Varda},
  {Sinyakov}, {Gabovich}, {Krushinsky}, {Zalozhnih}, {Popov}, {Bourdanov}, \&
  {Shurpakov}}]{2015ATel.6951....1S}
{Shumkov}, V., {Pruzhinskaya}, M., {Tiurina}, N., {et~al.} 2015{\natexlab{a}},
  ATel, 6951

\bibitem[{{Shumkov} {et~al.}(2015{\natexlab{b}}){Shumkov}, {Lipunov},
  {Gorbovskoy}, {Tiurina}, {Balanutsa}, {Kuznetsov}, {Pruzhinskaya},
  {Denisenko}, {Rufanov}, {Vladimirov}, {Ivanov}, {Yazev}, {Budnev}, {Gress},
  {Poleshchuk}, {Parkhomenko}, {Tlatov}, {Dormidontov}, {Yurkov}, {Sergienko},
  {Varda}, {Sinyakov}, {Gabovich}, {Krushinsky}, {Zalozhnih}, {Popov},
  {Bourdanov}, {Buckley}, {Potter}, {Kniazev}, {Kotze}, {Shumkov},
  {Vladimirov}, {Lipunov}, {Gorbovskoy}, {Tiurina}, {Balanutsa}, {Kuznetsov},
  {Pruzhinskaya}, {Denisenko}, {Rufanov}, {Ivanov}, {Yazev}, {Budnev}, {Gress},
  {Poleshchuk}, {Parkhomenko}, {Tlatov}, {Dormidontov}, {Yurkov}, {Sergienko},
  {Varda}, {Sinyakov}, {Gabovich}, {Krushinsky}, {Zalozhnih}, {Popov},
  {Bourdanov}, {Buckley}, {Potter}, {Kniazev}, {Kotze}, \&
  {Shurpakov}}]{2015ATel.6911....1S}
{Shumkov}, V., {Lipunov}, V., {Gorbovskoy}, E., {et~al.} 2015{\natexlab{b}},
  ATel, 6911

\bibitem[{{Sirianni} {et~al.}(2005){Sirianni}, {Jee}, {Ben{\'{\i}}tez},
  {Blakeslee}, {Martel}, {Meurer}, {Clampin}, {De Marchi}, {Ford}, {Gilliland},
  {Hartig}, {Illingworth}, {Mack}, \& {McCann}}]{2005PASP..117.1049S}
{Sirianni}, M., {Jee}, M.~J., {Ben{\'{\i}}tez}, N., {et~al.} 2005, \pasp, 117,
  1049

\bibitem[{{Soker} \& {Tylenda}(2003)}]{2003ApJ...582L.105S}
{Soker}, N., \& {Tylenda}, R. 2003, \apjl, 582, L105

\bibitem[{{Sparks} {et~al.}(2008){Sparks}, {Bond}, {Cracraft}, {Levay},
  {Crause}, {Dopita}, {Henden}, {Munari}, {Panagia}, {Starrfield}, {Sugerman},
  {Wagner}, \& {White}}]{2008AJ....135..605S}
{Sparks}, W.~B., {Bond}, H.~E., {Cracraft}, M., {et~al.} 2008, \aj, 135, 605

\bibitem[{{Steele} {et~al.}(2004){Steele}, {Smith}, {Rees}, {Baker}, {Bates},
  {Bode}, {Bowman}, {Carter}, {Etherton}, {Ford}, {Fraser}, {Gomboc}, {Lett},
  {Mansfield}, {Marchant}, {Medrano-Cerda}, {Mottram}, {Raback}, {Scott},
  {Tomlinson}, \& {Zamanov}}]{2004SPIE.5489..679S}
{Steele}, I.~A., {Smith}, R.~J., {Rees}, P.~C., {et~al.} 2004, SPIE, 5489, 679

\bibitem[{{Stone}(1977)}]{1977ApJ...218..767S}
{Stone}, R.~P.~S. 1977, \apj, 218, 767

\bibitem[{{Thompson} {et~al.}(2009){Thompson}, {Prieto}, {Stanek}, {Kistler},
  {Beacom}, \& {Kochanek}}]{2009ApJ...705.1364T}
{Thompson}, T.~A., {Prieto}, J.~L., {Stanek}, K.~Z., {et~al.} 2009, \apj, 705,
  1364

\bibitem[{{Tody}(1986)}]{1986SPIE..627..733T}
{Tody}, D. 1986, SPIE, 627, 733

\bibitem[{{Tylenda} {et~al.}(2011){Tylenda}, {Hajduk}, {Kami{\'n}ski},
  {Udalski}, {Soszy{\'n}ski}, {Szyma{\'n}ski}, {Kubiak}, {Pietrzy{\'n}ski},
  {Poleski}, {Wyrzykowski}, \& {Ulaczyk}}]{2011A&A...528A.114T}
{Tylenda}, R., {Hajduk}, M., {Kami{\'n}ski}, T., {et~al.} 2011, \aap, 528, A114

\bibitem[Wagner et al.(2015)]{2015ATel.7208....1W} Wagner, R.~M., 
Starrfield, S.~G., Wilber, A., et al.\ 2015, ATel, 
7208

\bibitem[{{Williams} {et~al.}(2014){Williams}, {Darnley}, {Bode}, {Keen}, \&
  {Shafter}}]{2014ApJS..213...10W}
{Williams}, S.~C., {Darnley}, M.~J., {Bode}, M.~F., {Keen}, A., \& {Shafter},
  A.~W. 2014, \apjs, 213, 10
  
  }
\end{thebibliography}

\end{document}